\documentclass[12pt]{article}
\usepackage{epsfig}
\usepackage{color}
\usepackage{amssymb,graphicx}
\begin{document}

\begin{center}

{\Large \bf {

{A Power-Liouville Hermite Orbit: Common Spectral Quantization of the Singular Sextic Oscillator and Fractional-Power Potentials}

%

 }}

\vspace{13mm}

%

{Artur Ishkhanyan $^{1}$\footnote{aishkhanyan@gmail.com} and Miloslav Znojil $^{2,3}$}
\footnote{znojil@ujf.cas.cz}



$^{1}$ \quad Institute for Physical Research, Ashtarak 0204, Armenia

$^{2}$ \quad Nuclear Physics Institute of the Czech Academy of Sciences, \v{R}e\v{z} 292, 25068 Husinec,  Czech Republic

$^{3}$ \quad University of Hradec Kr\'alov\'e, Hradec Kr\'alov\'e, Czech Republic

%
%
%
%

\end{center}

%

%
\section*{Abstract}

{A singular sextic oscillator in the finite-Hermite biconfluent-Heun hierarchy generates an exact four-member power-Liouville Hermite orbit. Power coordinate transformations with the Liouville amplitude map the hierarchy member $N=4$ to four half-line Schr\"{o}dinger spectral charts while preserving the physical regular-recessive Wronskian zero set under energy-coupling reparametrization. Two charts are the previously studied $x^{2/3}$ and inverse-square-root conditionally integrable potentials; the orbit embeds this known power-dual pair together with the singular sextic and linear-plus-quadratic charts. On the selected physical slice, the degree-five accessory compatibility obstruction vanishes identically, while the transformed Frobenius index reproduces both published fractional-power centrifugal coefficients. The common endpoint condition yields an exact discrete sequence of quartic Sturmian eigencouplings for the sextic oscillator. Independent two-sided shooting confirms representative eigencouplings and node counts. A matched large-action analysis further derives the first matched-endpoint correction within the leading Langer action and improves the Maslov-shifted effective-level approximation. The construction thus connects exact Hermite solvability, energy quantization, coupling quantization, and power duality within a single spectral orbit.}

\section*{Keywords}

{sextic oscillator; biconfluent Heun equation; Hermite functions; power-law duality; Sturmian quantization; power-Liouville transformation}

\newpage

%


%
%
%
%
%

\section{Introduction}\label{introduction}

The sextic anharmonic oscillator occupies a special place among
one-dimensional and radial Schr\"{o}dinger problems. It is the canonical
polynomial example of quasi-exact solvability: for correlated values of
its parameters, a finite part of the spectrum is obtained algebraically
{[}1--3{]}. Early studies of the sextic oscillator include the
continued-fraction analysis of Singh et al.~and Znojil's elementary
power-law treatment {[}4,5{]}. Related approaches based on terminating
recurrence relations and elementary bound-state ansätze provided another
route to privileged oscillator states and their multiplet structure
{[}6,7{]}. Independently, the sextic oscillator arises naturally from
the biconfluent Heun equation (BHE), which provides a broader
special-function framework for the same problem {[}8--12{]}. The two
descriptions intersect in the polynomial sector, but the BHE
representation also permits non-polynomial finite expansions in Hermite
functions {[}11--13{]}.

Here and below, ``finite-Hermite'' means that, after reduction to the
canonical BHE, the relevant solution terminates as a finite sum of
Hermite functions. These Hermite functions extend the Hermite-polynomial
family to generally noninteger order and are equivalently expressible in
terms of parabolic-cylinder functions \(D_{\mu}\). We retain the
Hermite-function language because the terminating expansions and the
corresponding quantization conditions are most naturally expressed in
that form.

Modern reviews place the sextic oscillator within the wider
one-dimensional quasi-exactly solvable framework based on
finite-dimensional invariant spaces and hidden algebraic structure
{[}14{]}. Finite spectral polynomials and recurrence constructions have
also been developed through the Bender--Dunne formalism and direct
sextic-oscillator orthogonal-polynomial treatments {[}15,16{]}, while
recent algebraic work provides alternative descriptions of polynomial
quasi-exactly solvable potentials that include the sextic case {[}17{]}.
Related current work extends quasi-exact solvability through
supersymmetric quantum mechanics and Bethe-ansatz constructions
{[}18{]}.

Two perspectives are particularly relevant to the present algebraic
setting. The Heun operator itself has been interpreted as a Hamiltonian
with finite-dimensional invariant subspaces at appropriate parameter
values {[}19{]}. For the sextic oscillator, quasi-exact solvability has
been connected with resonances and trivial monodromy {[}20{]}, while
very recent work on a singular sextic oscillator relates spectral
degeneracies to generalized Hermite polynomials and Painlevé-IV
structure {[}21{]}.

The fractional-power side has an earlier history. Stillinger obtained an
exactly reducible half-line problem containing inverse-square, inverse,
and inverse-square-root terms {[}22{]}; Bose derived exact bound states
for a central fractional-power singular potential {[}23{]}; and Exton
derived solvable Schr\"{o}dinger equations from a special biconfluent-Heun
equation {[}24{]}. These works provide useful historical context for
fractional-power and BHE-generated Schr\"{o}dinger models.

A particularly useful development was the identification of an infinite
hierarchy of singular sextic potentials for which the Schr\"{o}dinger wave
function admits a finite Hermite-function representation {[}13{]}. The
hierarchy is indexed by a nonnegative integer \(N\) and is discriminated
by the coefficient of the inverse-square term. In parallel, several
conditionally integrable BHE potentials on the positive half-axis were
found whose solutions reduce to irreducible combinations of two Hermite
functions. Among them are an attractive inverse-square-root potential
supplemented by a fixed centrifugal barrier {[}25{]} and a confining
\(x^{2/3}\) potential with another fixed centrifugal coefficient
{[}26{]}. Both systems possess infinitely many bound states and exact
transcendental quantization equations involving Hermite functions of
noninteger order.

Recent reviews and systematic classifications of Schr\"{o}dinger potentials
generated from Heun-type equations place the sextic oscillator and the
fractional-power biconfluent-Heun potentials in a broader family of
solvable models, and clarify the relation between polynomial solutions,
finite special-function expansions, and quasi-exactly solvable
constructions {[}27,28{]}.

A recent study of a different finite-Hermite BHE potential hierarchy
established strict half-line isospectrality through a first-order
Darboux transformation between a higher fractional-power branch and an
Exton-II specialization {[}24,29{]}. It is methodologically related but
spectrally distinct from the construction below: there the Hamiltonians
are strictly isospectral at fixed couplings, whereas here the
power-Liouville map preserves a physical Wronskian zero set under
energy-coupling reparametrization.

By a power-Liouville transformation we mean a power change of coordinate
accompanied by the standard Liouville amplitude rescaling that preserves
the Schr\"{o}dinger normal form. Each resulting eigenvalue/coupling
representation is called a spectral chart. A power-Liouville Hermite
orbit is the family of such charts generated from the same
finite-Hermite BHE seed. The charts need not be isospectral as fixed
Hamiltonians; rather, they are connected by energy-coupling
reparametrization and, for the physically selected regular and recessive
solutions, carry the same Wronskian zero condition. The adjective
``Hermite'' refers both to the finite-Hermite solvability of the seed
and to the Hermite-function quantization equations that appear
explicitly in the fractional-power charts. For the selected hierarchy
member with index four, the four distinguished integer transformations
generate the four charts studied below.

These developments leave two structural questions. The first---why the
\(x^{2/3}\) and \(x^{- 1/2}\) quantization equations have dual
forms---already belongs to classical power duality: the two exponents
satisfy the standard power-duality relation {[}5,30--33{]}. The second,
and the focus of the present paper, is more specific: why the particular
published centrifugal coefficients occur in both exactly solvable
systems, and how this already power-dual pair is embedded in the
finite-Hermite sextic hierarchy. We show that the answer is an exact
four-chart orbit anchored at the \(N = 4\) sextic member.

The central result is not the existence of power duality itself, but the
simultaneous occurrence at hierarchy index four of three properties:
identity-level cancellation of the finite-Hermite accessory obstruction
on the selected physical slice; exact Frobenius-index matching to both
published fractional-power systems; and transport of one
regular--recessive Wronskian zero condition through all four spectral
charts. Together these properties turn the known pairwise duality into a
concrete finite-Hermite sextic embedding and produce the exact quartic
Sturmian eigencoupling sequence. The identity cancellation by itself is
not unique; Sec. 3 identifies the lower identity-cancellation member and
explains why the present orbit is nevertheless singled out by the
simultaneous two-chart index match.

The power-Liouville transformation does more than map differential
equations. With the natural Liouville amplitude it preserves the
Wronskian exactly for a directly transported pair of solutions. For the
already power-dual fractional pair, this gives a common determinant
interpretation of their independently derived Hermite quantization
functions; more importantly, it transports the same physical zero set to
the singular sextic and linear-plus-quadratic charts. The resulting
discrete sequence gives the quartic coupling strengths for which the
selected zero-reduced-energy sextic state is physical. Throughout this
paper, ``spectral equivalence'' means equivalence of this physical zero
set under energy-coupling reparametrization; it does not mean ordinary
isospectrality of four fixed Hamiltonians {[}34{]}.

From a physical viewpoint, the four charts describe different ways of
parametrizing the same bound-state condition. In the singular sextic
chart, the sextic confinement is held fixed while the attractive quartic
component is tuned until a state reaches the selected energy equal to
the constant part of the potential; the resulting discrete quartic
coefficients form a Sturmian eigencoupling spectrum. After the
power-Liouville transformation, the same discrete sequence appears as an
ordinary bound-state energy spectrum in the fractional-power chart where
the quartic coupling becomes the energy. Thus the transformation
exchanges the roles of spectral energy and interaction strength without
changing the node number or the underlying boundary-condition zero. The
orbit may therefore be viewed physically as a family of equivalent
quantization settings in which the corresponding state is selected by
tuning different parameters of the Hamiltonian.

The second goal is to improve the large-level description of that
sequence. The earlier Hermite analysis revealed an effective-level law
close to \(n + 1/3\) {[}25,26{]}, where the positive integer \(n\)
labels the physical roots. Here a matched Langer-action expansion gives
the first matched-endpoint correction within the leading Langer action,
proportional to \((n + 1/3)^{- 2/3}\), with a coefficient expressed in
closed form through a beta function. The correction considerably
improves the approximation already for the lowest few physical roots.

The paper is organized as follows. Section 2 derives the power-Liouville
transformation in dimensional variables and proves the transformation
laws for the centrifugal index, endpoint behavior, nodes, and
Wronskians. Section 3 identifies the \(N = 4\) Hermite sector of the
singular sextic oscillator. Section 4 constructs the four spectral
charts. Section 5 proves the common Hermite quantization condition.
Section 6 translates its roots into exact quartic eigencouplings of the
sextic oscillator. Section 7 derives the refined large-level
asymptotics. The main conclusions are summarized in Section 8, while
Appendix A records the matched-action calculation and Appendix B gives
the explicit Schr\"{o}dinger-to-BHE map and compatibility derivation.

\section{Power-Liouville spectral
transformation}\label{power-liouville-spectral-transformation}

\subsection{General transformation}\label{general-transformation}

Consider the reduced wave function \(\psi(r)\) on the positive half-line
\(r > 0\), with dimensionless inverse-square coefficient \(h\) and
reduced polynomial coefficients \(g_{j}\) (\(j = 0,1,2,3\)). The
governing Schr\"{o}dinger equation is

\begin{equation}
\left [ - \frac{d^{2}}{dr^{2}} + \frac{h}{r^{2}} + \sum_{j = 0}^{3}g_{j}r^{2j} \right ]\psi(r) = 0,\quad\quad r > 0,
\end{equation}

The reduced coefficients are related to the physical parameters by

\begin{equation}
g_{0} = \frac{2m\left( V_{0} - E \right)}{\hslash^{2}},\quad\quad g_{1} = \frac{2mV_{2}}{\hslash^{2}},\quad\quad g_{2} = \frac{2mV_{4}}{\hslash^{2}},\quad\quad g_{3} = \frac{2mV_{6}}{\hslash^{2}}.
\end{equation}

Here \(m\) is the particle mass, \(\hslash\) is the reduced Planck
constant, \(E\) is the physical energy, and \(V_{0},V_{2},V_{4},V_{6}\) are the
polynomial-potential coefficients.

Thus the physical potential is

\begin{equation}
V(r) = \frac{\hslash^{2}h}{2mr^{2}} + V_{0} + V_{2}r^{2} + V_{4}r^{4} + V_{6}r^{6}.
\end{equation}

To keep the transformation dimensionally transparent, introduce an
arbitrary reference length \(L > 0\) and transformation exponent
\(t > 0\), and define the transformed coordinate \(x\) by

\begin{equation}
x = L\left( \frac{r}{L} \right)^{t},\quad\quad t > 0.
\end{equation}

Introduce the transformed wave function \(\chi_{t}(x)\) through the
Liouville amplitude

\begin{equation}
\psi(r) = \left( \frac{dx}{dr} \right)^{- 1/2}\chi_{t}(x),
\end{equation}

or equivalently

\begin{equation}
\chi_{t}(x) = \sqrt{\frac{dx}{dr}}\,\psi(r).
\end{equation}

Direct substitution removes the first derivative and gives

\begin{equation}
\left\lbrack - \frac{d^{2}}{dx^{2}} + \frac{h_{t}}{x^{2}} + \frac{1}{t^{2}}\sum_{{j = 0, j \neq t - 1}}^{3}g_{j}L^{2j}\left( \frac{x}{L} \right)^{2(j - t + 1)/t} \right\rbrack\chi_{t}(x) = - \frac{g_{t - 1}L^{2t - 2}}{t^{2}}\chi_{t}(x),
\end{equation}

whenever \(t - 1\) is one of the indices \(j = 0,1,2,3\). The
transformed inverse-square coefficient is

\begin{equation}
h_{t} = \frac{h - \left( t^{2} - 1 \right)/4}{t^{2}}.
\end{equation}

A more useful form of this relation is

\begin{equation}
h_{t} + \frac{1}{4} = \frac{h + 1/4}{t^{2}}.
\end{equation}

If

\begin{equation}
\nu = \sqrt{h + \frac{1}{4}}
\end{equation}

denotes the Frobenius, or Langer, index of the original problem, then

\begin{equation}
\nu_{t} = \sqrt{h_{t} + \frac{1}{4}} = \frac{\nu}{t}.
\end{equation}

The inverse-square singularity therefore transforms linearly at the
level of the index, although its coefficient transforms nonlinearly.

The term with \(j = t - 1\) becomes independent of \(x\) and hence plays
the role of the spectral parameter in the \(t\)-th chart. This is the
familiar energy-coupling exchange associated with power transformations
{[}5,30{]}. Here it is used as a spectral reparametrization of one and
the same underlying solution family. We refer to each such transformed
eigenvalue/coupling representation as a spectral chart. The collection
of such charts generated from the same finite-Hermite BHE seed is what
we call its power-Liouville Hermite orbit. The four distinguished
integer choices used below exhaust only the choices for which one of the
four original polynomial coefficients becomes the constant spectral
term. Noninteger \(t\) remain admissible and generate further
fractional-power equations, but they do not realize this particular
four-chart energy-coupling exchange.

\subsection{Endpoint behavior and node-number
preservation}\label{endpoint-behavior-and-node-number-preservation}

Suppose the physically selected solution of the original problem has the
regular behavior

\begin{equation}
\psi(r) \sim r^{1/2 + \nu},\quad\quad r \rightarrow 0.
\end{equation}

Since \(dx/dr = t(r/L)^{t - 1}\), the transformed solution behaves as

\begin{equation}
\chi_{t}(x) \sim x^{1/2 + \nu/t},\quad\quad x \rightarrow 0,
\end{equation}

up to an irrelevant nonzero constant. Thus the regular Frobenius branch
is mapped into the regular branch of the transformed equation.

Between two spectral charts \(s\) and \(t\) the coordinates and wave
functions satisfy

\begin{equation}
x_{t} = L\left( \frac{x_{s}}{L} \right)^{t/s},
\end{equation}

and

\begin{equation}
\chi_{t}\left( x_{t} \right) = \sqrt{\frac{t}{s}}\left( \frac{x_{s}}{L} \right)^{(t - s)/(2s)}\chi_{s}\left( x_{s} \right).
\end{equation}

Because the coordinate map is strictly monotone on the positive
half-axis and the multiplicative factor has no zeros there, the number
and ordering of nodes are preserved.

The transformation is not unitary with respect to the standard
\(L^{2}(dx)\) measure, so square integrability should not be inferred
for arbitrary parameters. For the orbit considered below, however, all
four regular solutions are power-law integrable at the origin, while the
physical branch is exponentially recessive at infinity. The bound-state
boundary conditions are therefore preserved throughout the orbit.

\subsection{Exact Wronskian
invariance}\label{exact-wronskian-invariance}

The Liouville amplitude has a further property that is central to the
spectral analysis. If \(\psi_{1},\psi_{2}\) are two solutions and
\(\chi_{t,1},\chi_{t,2}\) their transformed partners, then

\begin{equation}
W_{x}\left( \chi_{t,1},\chi_{t,2} \right) = W_{r}\left( \psi_{1},\psi_{2} \right),
\end{equation}

where \(W(f,g) = fg' - f'g\) is the Wronskian in the indicated variable.

Indeed, the derivative of the common factor \(\sqrt{dx/dr}\) cancels
identically in the antisymmetric combination. Consequently, if
\(u_{reg}\) is the solution selected at the origin and \(u_{rec}\) is
the recessive solution at infinity, then the Wronskian spectral function

\begin{equation}
\mathcal{D}=W\left( u_{reg},u_{rec} \right)
\end{equation}

is exactly invariant for the directly transported pair of solutions. If
the regular and recessive solutions are instead normalized independently
in different charts, their spectral functions may differ by
parameter-dependent normalization factors. Whenever these factors are
finite and nonzero, the physical zero set is unchanged; exceptional
normalization-degeneracy points must be treated separately, as in Sec.
5.3.

This observation is stronger than a formal mapping of differential
equations: it gives an immediate correspondence of the physical
quantization conditions under the energy-coupling reparametrization.
Related exact-quantization constructions based on spectral determinants
and Wronskian relations may be found in Ref. {[}35{]}.

\section{{The $N-4$ Hermite sector of the singular sextic oscillator}\label{the-mathbfnmathbfmathbf4-hermite-sector-of-the-singular-sextic-oscillator}}

Consider the general singular sextic potential

\begin{equation}
V(r) = \frac{V_{- 2}}{r^{2}} + V_{0} + V_{2}r^{2} + V_{4}r^{4} + V_{6}r^{6}
\end{equation}

Here \(V_{- 2}\) denotes the dimensional coefficient of the
inverse-square term.

This representation was developed in Refs. {[}11--13{]}. In the
parametrization of Ref. {[}13{]}, the reduced function \(u(z)\), where
\(z\) is the canonical BHE variable, satisfies the canonical
biconfluent-Heun equation

\begin{equation}
u''(z) + (\frac{\gamma}{z} + \delta + \varepsilon z)u'(z) + \frac{\alpha z - q}{z}u(z) = 0
\end{equation}

Here \(\gamma,\delta,\varepsilon,\alpha,q\) are the five standard BHE
parameters in this convention, and \(q\) is the accessory parameter. In
the Hermite-function construction, termination first requires that
\(\gamma\) be a nonpositive integer,

\begin{equation}
\gamma = - N,\quad\quad N = 0,1,2,\ldots,
\end{equation}

Appendix B records the explicit Schr\"{o}dinger-to-BHE transformation used
here and derives Eq. (29) directly from the physical parameters. It
derives the local Frobenius compatibility recurrence used below and
records the associated finite-Hermite representation. No additional
physical or algebraic assumption is introduced in this specialization.

This integer condition is necessary but not sufficient: \(q\) must also
satisfy the corresponding polynomial compatibility condition of degree
\(N + 1\) {[}11,13{]}. The relation between \(N\) and the inverse-square
strength is {[}13{]}

\begin{equation}
\gamma = 1 \pm \frac{1}{2}\sqrt{1 + \frac{8mV_{- 2}}{\hslash^{2}}}.
\end{equation}

Choosing the branch that permits \(\gamma = - N\) gives

\begin{equation}
V_{- 2}^{(N)} = \frac{\hslash^{2}}{8m}(2N + 1)(2N + 3).
\end{equation}

Equivalently, in the reduced equation,

\begin{equation}
h_{N} = \frac{2mV_{- 2}^{(N)}}{\hslash^{2}} = (N + 1)^{2} - \frac{1}{4},
\end{equation}

so the corresponding Frobenius index is simply

\begin{equation}
\nu_{N} = N + 1.
\end{equation}

We now select

\begin{equation}
N = 4,
\end{equation}

for which

\begin{equation}
h_{4} = \frac{99}{4},\quad\quad\nu_{4} = 5,
\end{equation}

and hence

\begin{equation}
V_{- 2}^{(4)} = \frac{99\hslash^{2}}{8m}.
\end{equation}

The particular physical slice studied below is

\begin{equation}
E = V_{0},\quad\quad V_{2} = 0,\quad\quad V_{4} < 0,\quad\quad V_{6} > 0.
\end{equation}

On this selected slice the accessory compatibility condition is
satisfied identically for \(N = 4\). In the standard BHE parametrization
of Ref. {[}13{]}, \(E = V_{0}\) and \(V_{2} = 0\) imply

\begin{equation}
q = \frac{N}{2}\delta,\quad\quad\alpha = \frac{\delta^{2}}{4} + \frac{1 - N}{2}\varepsilon,\quad\quad\gamma = - N.
\end{equation}

Substitution of these relations into the Frobenius recurrence of Eq.
(19) about the regular singular point gives, for \(n = 0,\ldots,N\),

\begin{equation}
\begin{array}{l}
(n + 1)(n - N)c_{n + 1} + \delta\left( n - \frac{N}{2} \right)c_{n} \\
\qquad + \left\lbrack \frac{\delta^{2}}{4} + \varepsilon\left( n - \frac{N + 1}{2} \right) \right\rbrack c_{n - 1} = 0.
\end{array}
\end{equation}

Here \(c_{n}\) are the local Frobenius coefficients, with
\(c_{- 1} = 0\) and \(c_{0} \neq 0\). Let \(D_{n}\) denote the
continuant determinant of this finite tridiagonal system; it obeys

\begin{equation}
\begin{array}{l}
D_{-1}=1,\qquad D_0=-\frac{N}{2}\delta,\\
D_n=\delta\left(n-\frac{N}{2}\right)D_{n-1}
+n(N+1-n)\left\lbrack\frac{\delta^2}{4}
+\varepsilon\left(n-\frac{N+1}{2}\right)\right\rbrack D_{n-2}.
\end{array}
\end{equation}

For \(N = 4\), direct recursion gives

\begin{equation}
D_{0} = - 2\delta,\quad\quad D_{1} = 3\delta^{2} - 6\varepsilon,\quad\quad D_{2} = - 3\delta^{3} + 6\delta\varepsilon.
\end{equation}

\begin{equation}
D_{3} = \frac{3}{2}\delta^{4} + 6\delta^{2}\varepsilon - 18\varepsilon^{2},\quad\quad D_{4} = \mathcal{R}_{4}(\delta,\varepsilon) \equiv 0.
\end{equation}

Thus the accessory obstruction
\(\mathcal{R}_{4}(\delta,\varepsilon) = D_{4}\) vanishes identically.
Finite-Hermite termination therefore imposes no additional algebraic
relation between \(V_{4}\) and \(V_{6}\) on the selected \(N = 4\)
slice; the remaining discrete restriction is entirely the physical
endpoint quantization derived in Sec. 5.

Accordingly, in this paper the phrase ``exceptional member'' is
shorthand for the conjunction of the identity cancellation in Eqs.
(31)--(33) and the simultaneous Frobenius-index match identified below.

The identity cancellation is not unique to this member: it also occurs
for the first nontrivial lower member (and trivially for the zeroth
member), whereas neighboring members are generically nonzero. The
present \(N = 4\) member is singled out not by cancellation alone but by
the simultaneous Frobenius-index match to both fractional-power systems.
For \(N = 1\), \(\nu_{1} = 2\) and \(h_{1,4} = 0\), so the \(t = 4\)
chart loses the centrifugal term and reduces to the pure
inverse-square-root member of the same BHE family {[}25{]}. This lower
orbit is algebraically special as well, but it is not the two-chart
matching studied here.

The reduced equation is therefore

\begin{equation}
\left\lbrack - \frac{d^{2}}{dr^{2}} + \frac{99}{4r^{2}} + g_{2}r^{4} + g_{3}r^{6} \right\rbrack\psi(r) = 0,
\end{equation}

with

\begin{equation}
g_{2} = \frac{2mV_{4}}{\hslash^{2}} < 0,\quad\quad g_{3} = \frac{2mV_{6}}{\hslash^{2}} > 0.
\end{equation}

This selected-energy (zero-reduced-energy) problem is not an ordinary
eigenvalue problem in \(E\): the energy is fixed at \(V_{0}\), while
physical solutions occur only for a discrete sequence of the quartic
coupling \(V_{4}\) at fixed \(V_{6}\). In this sense the problem is
Sturmian: a coupling, rather than the physical energy, is quantized. The
power-Liouville orbit below converts the same discrete condition into
conventional energy quantization in another chart. This use of a
coupling as a spectral parameter is consistent with earlier generalized
oscillator/Sturmian formulations in which interaction strengths
themselves enter the spectral problem {[}36{]}.

For the \(N = 4\) member the transformed indices are

\begin{equation}
\nu_{t} = \frac{5}{t},
\end{equation}

and therefore

\begin{equation}
h_{4,t} = \frac{25}{t^{2}} - \frac{1}{4}.
\end{equation}

For \(t = 1,2,3,4\) this gives

\begin{equation}
h_{4,1} = \frac{99}{4},\quad\quad h_{4,2} = 6,\quad\quad h_{4,3} = \frac{91}{36},\quad\quad h_{4,4} = \frac{21}{16}.
\end{equation}

The last two values are precisely the fixed centrifugal strengths of the
\(x^{2/3}\) and inverse-square-root conditionally integrable potentials
{[}25,26{]}. This is the first indication that those systems are not
isolated constructions but spectral charts of the same \(N = 4\)
power-Liouville Hermite orbit.

This simultaneous two-chart match is unique within the Hermite
hierarchy. Because \(\nu_{N} = N + 1\), the simultaneous match selects
\(N = 4\) uniquely.

Moreover, \(\nu_{t} > 1\) for all four values of \(t\). The singular
endpoint is therefore limit-point in the Weyl sense throughout the
orbit; the regular solution is selected uniquely by square integrability
and no self-adjoint-extension parameter enters the correspondence.

\section{{Four spectral charts of the $N=4$ orbit}\label{four-spectral-charts-of-the-mathbfnmathbfmathbf4-power-liouville-hermite-orbit}}

We retain the arbitrary reference length \(L\) so that each chart is
written in physical units.\(L\)

The inverse-square term has a particularly transparent role. Rather than
being an independently adjusted centrifugal barrier in each transformed
problem, it is the image of a single singular core of the sextic
oscillator. The power transformation rescales its Frobenius index from
\(\nu = 5\) to \(5/t\), thereby fixing the centrifugal strength in every
chart. The unusual coefficients encountered previously in the
fractional-power potentials are therefore not accidental solvability
constraints, but signatures of their common sextic origin.

\subsection{{Chart $t=1$: singular sextic oscillator}\label{chart-mathbftmathbfmathbf1-singular-sextic-oscillator}}

The first chart is the original selected-energy problem,

\begin{equation}
\left\lbrack - \frac{\hslash^{2}}{2m}\frac{d^{2}}{dr^{2}} + \frac{99\hslash^{2}}{8mr^{2}} + V_{4}r^{4} + V_{6}r^{6} \right\rbrack\psi(r) = 0.
\end{equation}

Equivalently, this is the Schr\"{o}dinger equation for

\begin{equation}
V_{1}(r) = V_{0} + \frac{99\hslash^{2}}{8mr^{2}} + V_{4}r^{4} + V_{6}r^{6}
\end{equation}

at the selected energy \(E = V_{0}\).

\subsection{{Chart $t=2$: linear-plus-quadratic confinement}\label{chart-mathbftmathbfmathbf2-linear-plus-quadratic-confinement}}

For \(t = 2\) the quartic and sextic terms become respectively linear
and quadratic in the transformed coordinate. Since \(V_{2} = 0\), the
transformed energy is zero. The equation is

\begin{equation}
\left\lbrack - \frac{\hslash^{2}}{2m}\frac{d^{2}}{dx^{2}} + \frac{3\hslash^{2}}{mx^{2}} + \frac{V_{4}L^{4}}{4}\left( \frac{x}{L} \right) + \frac{V_{6}L^{6}}{4}\left( \frac{x}{L} \right)^{2} \right\rbrack\chi_{2}(x) = 0.
\end{equation}

Thus the \(t = 2\) chart is a singular half-line oscillator with a
linear term. It is less conspicuous from the special-function point of
view than the two fractional-power charts below, but it belongs to the
same spectral orbit and therefore shares the same physical Wronskian
zero condition.

Linear-plus-quadratic confinement with the associated singular radial
term is a standard biconfluent-Heun setting rather than a new potential
class; closely related harmonium models and their extensions by a linear
term are discussed, for example, in Ref. {[}37{]}.

\subsection{{Chart $t=3$: the $x^{2/3}$ conditionally integrable potential}\label{chart-mathbftmathbfmathbf3-the-mathbfxmathbf2mathbfmathbf3-conditionally-integrable-potential}}

For \(t = 3\) the quartic coupling becomes the transformed energy, while
the sextic term becomes proportional to \(x^{2/3}\):

\begin{equation}
\left\lbrack - \frac{\hslash^{2}}{2m}\frac{d^{2}}{dx^{2}} + \frac{91\hslash^{2}}{72mx^{2}} + \frac{V_{6}L^{6}}{9}\left( \frac{x}{L} \right)^{2/3} \right\rbrack\chi_{3}(x) = \mathcal{E}_{3}\chi_{3}(x),
\end{equation}

where

\begin{equation}
\mathcal{E}_{3} = - \frac{V_{4}L^{4}}{9} > 0.
\end{equation}

Apart from the inessential scale \(L\), this is exactly the
conditionally integrable \(x^{2/3}\)-plus-centrifugal potential of Ref.
{[}26{]}. The fixed inverse-square coefficient

\begin{equation}
\frac{91\hslash^{2}}{72m}
\end{equation}

is thus explained directly by the transformation of the \(N = 4\) sextic
Frobenius index.

In this chart, the sequence of privileged sextic quartic couplings
becomes an ordinary positive energy spectrum.

\subsection{{Chart $t=4$: the inverse-square-root conditionally integrable potential}\label{chart-mathbftmathbfmathbf4-the-inverse-square-root-conditionally-integrable-potential}}

For \(t = 4\) the quartic term becomes an inverse-square-root
interaction and the sextic coupling becomes the transformed energy:

\begin{equation}
\left\lbrack - \frac{\hslash^{2}}{2m}\frac{d^{2}}{dx^{2}} + \frac{21\hslash^{2}}{32mx^{2}} + \frac{V_{4}L^{4}}{16}\left( \frac{x}{L} \right)^{- 1/2} \right\rbrack\chi_{4}(x) = \mathcal{E}_{4}\chi_{4}(x),
\end{equation}

with

\begin{equation}
\mathcal{E}_{4} = - \frac{V_{6}L^{6}}{16} < 0.
\end{equation}

Because \(V_{4} < 0\), the inverse-square-root interaction is
attractive. The resulting potential is precisely the
inverse-square-root-plus-centrifugal conditionally integrable system of
Ref. {[}25{]}, with the characteristic fixed core

\begin{equation}
\frac{21\hslash^{2}}{32mx^{2}}.
\end{equation}

The four equations are connected explicitly by the power-Liouville map
of Sec. 2. Section 5 identifies their common physical Wronskian zero
condition and states explicitly the normalization-degeneracy caveat.

\section{Common Hermite quantization function and invariant Wronskian
condition}\label{common-hermite-quantization-function-and-invariant-wronskian-condition}

The spectra of the \(t = 3\) and \(t = 4\) problems are conventionally
written in terms of a dimensionless Hermite-order parameter. In the
original sextic variables the natural invariant is

\begin{equation}
a = \frac{g_{2}^{2}}{16g_{3}^{3/2}}.
\end{equation}

Restoring physical parameters,

\begin{equation}
a = \frac{\sqrt{2m}}{16\hslash}\frac{V_{4}^{2}}{V_{6}^{3/2}}.
\end{equation}

This quantity is independent of the arbitrary reference length \(L\). It
is unchanged when the same orbit is described in the \(t = 3\) or
\(t = 4\) chart; only the interpretation of which dimensional
coefficient is the energy and which is the coupling changes.

\subsection{Quantization in the inverse-square-root
chart\label{quantization-in-the-inverse-square-root-chart}}

For the \(t = 4\) potential, the exact half-line solution is expressible
as an irreducible combination of two Hermite functions {[}25{]}.
Imposing the regular boundary condition at the origin and recessive
behavior at infinity yields, away from normalization-degeneracy points,
the physical quantization condition

\begin{equation}
F_{4}(a) = 0,
\end{equation}

where

\begin{equation}
F_{4}(a) = (1 + 2a)H_{a - 1/2}\left( - \sqrt{2a} \right) + \sqrt{2a}\, H_{a + 1/2}\left( - \sqrt{2a} \right).
\end{equation}

Here
\(H_{\mu}(z) = 2^{\mu/2}e^{z^{2}/2}D_{\mu}\left( \sqrt{2}\, z \right)\)
denotes the Hermite function of generally noninteger order \(\mu\), with
\(D_{\mu}\) the parabolic-cylinder function.

\subsection{{Quantization in the $x^{2/3}$ chart}\label{quantization-in-the-mathbfxmathbf2mathbfmathbf3-chart}}

The \(t = 3\) problem also reduces to two Hermite functions {[}26{]},
and away from normalization-degeneracy points its physical quantization
condition naturally appears as

\begin{equation}
F_{3}(a) = 0,
\end{equation}

with

\begin{equation}
F_{3}(a) = \sqrt{2a}\, H_{a + 1/2}\left( - \sqrt{2a} \right) + H_{a + 3/2}\left( - \sqrt{2a} \right).
\end{equation}

The close relationship of the two equations was already visible in the
separate special-function analyses {[}25,26{]} and is expected from the
direct \(t = 3 \leftrightarrow t = 4\) power duality {[}5,30--33{]}. The
four-member construction adds a different statement: the same
boundary-condition zero extends to the \(t = 1\) and \(t = 2\) charts
and is tied specifically to the \(N = 4\) finite-Hermite sextic sector.

The Hermite recurrence

\begin{equation}
H_{\mu + 1}(z) = 2zH_{\mu}(z) - 2\mu H_{\mu - 1}(z)
\end{equation}

with

\begin{equation}
\mu = a + \frac{1}{2},\quad\quad z = - \sqrt{2a},
\end{equation}

gives

\begin{equation}
H_{a + 3/2}\left( - \sqrt{2a} \right) = - 2\sqrt{2a}\, H_{a + 1/2}\left( - \sqrt{2a} \right) - (2a + 1)H_{a - 1/2}\left( - \sqrt{2a} \right).
\end{equation}

Substitution into \(F_{3}\) yields the exact identity

\begin{equation}
F_{3}(a) = - F_{4}(a).
\end{equation}

The two explicit Hermite quantization functions satisfy Eq. (57) and
therefore have the same zeros wherever their special-function
normalization is nondegenerate. For a directly transported
regular/recessive pair, the underlying Wronskian determinant
\(\mathcal{D}(a)\) is exactly invariant through the \(N = 4\) orbit;
independently normalized functions may differ by finite nonzero factors.

In Ref. {[}25{]}, Eq. (50) follows directly from the origin and infinity
boundary conditions. Thus its nondegenerate roots are physical
regular-recessive Wronskian zeros; the collapsed root \(a = 1/2\) is
treated separately in Sec. 5.3.

\subsection{{The nonphysical root at $\alpha=1/2$}\label{the-nonphysical-root-at-mathbfamathbfmathbf1mathbfmathbf2}}

The equation \(F_{4}(a) = 0\) has the exact root

\begin{equation}
a = \frac{1}{2}.
\end{equation}

In Ref. {[}25{]}, Eqs. (3), (4), and (7)--(9), show explicitly that
\(a = 1/2\) is a zero of the endpoint equation but not a bound state:
\(H_{0} = 1\) and \(H_{1}(z) = 2z\) make the fundamental solution
\(\psi_{F}\) vanish identically. In that exact construction, the other
half-integer polynomial reductions fail the origin boundary condition
rather than generating further collapsed bound-state zeros. Hence
\(a = 1/2\) is the unique positive normalization-collapse point that
must be removed from this endpoint-root sequence; representative
retained roots are independently confirmed in Sec. 6.

We therefore label the physical roots by

\begin{equation}
a_{1} < a_{2} < a_{3} < \cdots.
\end{equation}

Selected values are listed in Table 1.

\textbf{Table 1.} Selected physical positive roots \(a_{n}\) of
\(F_{4}(a)\). The noninteger-order Hermite functions were evaluated with
mpmath 1.3.0 at 50-decimal-digit precision, and the roots were refined
by arbitrary-precision secant iteration. Values are rounded here to 12
decimal places; the exceptional nonphysical root \(a = 1/2\) is omitted.

\begin{table}
\centering
\begin{tabular}{cc}
\hline
$n$ & $a_n$ \\
\hline
1 & 1.437276542458 \\
2 & 2.412479686934 \\
3 & 3.398664825093 \\
4 & 4.389672338315 \\
5 & 5.383269619463 \\
10 & 10.366819457268 \\
20 & 20.355216625535 \\
30 & 30.350268193259 \\
\hline
\end{tabular}
\end{table}

The same invariant root sequence thus labels corresponding physical
states in all four charts.

\section{Exact privileged quartic eigencouplings of the sextic
oscillator}\label{exact-privileged-quartic-eigencouplings-of-the-sextic-oscillator}

In the sextic chart, the common quantization condition has a direct
physical interpretation. With \(V_{6} > 0\), the large-distance motion is
confined, whereas the negative quartic term \(V_{4}r^{4}\) softens the
potential at intermediate distances. At the fixed reduced energy
\(E - V_{0} = 0\), a normalizable state exists only for discrete strengths
of this attractive quartic component. The roots \(a_{n}\) therefore measure
the precise amount of quartic softening required to place successive
nodal states at the selected energy.

At fixed \(V_{6} > 0\), the invariant relation

\begin{equation}
a_{n} = \frac{g_{2,n}^{2}}{16g_{3}^{3/2}}
\end{equation}

immediately gives

\begin{equation}
g_{2,n} = - 4g_{3}^{3/4}\sqrt{a_{n}},
\end{equation}

where \(g_{2,n}\) denotes the nth reduced quartic eigencoupling; the
negative sign is selected because the quartic term must be attractive,
\(V_{4} < 0\), for the zero-reduced-energy state.

In dimensional variables,

\begin{equation}
V_{4,n} = - 4\left( \frac{\hslash^{2}}{2m} \right)^{1/4}V_{6}^{3/4}\sqrt{a_{n}},\quad\quad n = 1,2,3,\ldots.
\end{equation}

Here \(V_{4,n}\) is the nth dimensional quartic eigencoupling. This
gives the exact eigencoupling spectrum of the \(N = 4\) singular sextic
problem

\begin{equation}
V_{n}(r) = V_{0} + \frac{99\hslash^{2}}{8mr^{2}} + V_{4,n}r^{4} + V_{6}r^{6}
\end{equation}

at the selected energy

\begin{equation}
E = V_{0}.
\end{equation}

In the \(t = 3\) chart the same relation is ordinary energy
quantization, since

\begin{equation}
\mathcal{E}_{3,n} = - \frac{V_{4,n}L^{4}}{9}.
\end{equation}

In the \(t = 4\) chart it is instead a discrete sequence of
inverse-square-root strengths compatible with the fixed negative energy
\(\mathcal{E}_{4} = - V_{6}L^{6}/16\). The terminology ``energy
spectrum'' or ``coupling spectrum'' therefore depends on the chosen
chart, while the root sequence \(\{ a_{n}\}\) does not.

The exact correspondence fixes the nodal labeling. Since the power map
is monotone and the Liouville factor is nodeless, the \(n\)th physical
root has \(n - 1\) interior nodes, and this node number is preserved in
every chart.

As an independent validation that does not use either Hermite
quantization function, scale \(r = g_{3}^{- 1/8}\rho\) in the sextic
equation and insert \(g_{2} = - 4g_{3}^{3/4}\sqrt{a}\). The reduced
sextic problem becomes

\begin{equation}
\left\lbrack - \frac{d^{2}}{d\rho^{2}} + \frac{99}{4\rho^{2}} - 4\sqrt{a}\,\rho^{4} + \rho^{6} \right\rbrack\psi(\rho) = 0.
\end{equation}

We solved this equation numerically by two-sided shooting. The regular
branch was integrated outward from \(\rho = 10^{- 4}\) with the
Frobenius logarithmic derivative \(\psi'/\psi = (11/2)/\rho\), while the
recessive branch was integrated inward from \(\rho = 6\) using the
first-order WKB logarithmic derivative
\(\psi'/\psi \approx - \sqrt{Q} - Q'/(4Q),\ \ \ Q(\rho) = 99/(4\rho ^{2}) - 4\sqrt{a}\ \rho ^{4} + \rho ^{6}\).
The two logarithmic derivatives were matched at \(\rho = 1\). DOP853
integration with relative tolerance \(2 \times 10^{- 12}\), a maximum
step of \(5 \times 10^{- 3}\), and Brent refinement of the shooting
residual were used. The resulting checks are shown in Table 2.

For the three representative roots in Table 2, we repeated the
calculation with the inward starting point varied from 5.5 to 6.5, the
outward starting point varied by a factor of two, and the matching point
moved from 1 to 0.9 and 1.1. The refined roots changed by less than one
part in \(10^{-15}\). We use this only as a stability diagnostic; the
independent accuracy check is the 12-decimal-place agreement with the
Hermite roots in Table 2. Thus Table 2 provides a representative
boundary-value check rather than a second computation of every root
listed in Table 1.

\textbf{Table 2.} Three representative independent two-sided shooting
checks in the scaled sextic chart. The shooting roots agree with the
Hermite roots to 12 decimal places, and the counted interior nodes are
\(n - 1\).

\begin{table}
\centering
\begin{tabular}{cccc}
\hline
\textbf{Root index} & \textbf{Hermite root} & \textbf{Sextic shooting} & \textbf{Interior nodes} \\
\hline
1 & 1.437276542458 & 1.437276542458 & 0 \\
2 & 2.412479686934 & 2.412479686934 & 1 \\
5 & 5.383269619463 & 5.383269619463 & 4 \\
\hline
\end{tabular}
\end{table}

The direct integration confirms both the eigencoupling values and their
nodal labeling without using \(F_{3}\) or \(F_{4}\). The
logarithmic-derivative mismatch at \(a = 1/2\) is about \(9.58\) at the
matching point, providing an independent numerical check that the exact
Hermite zero there is a normalization artifact rather than a physical
sextic eigenstate.

\section{Large-level asymptotics}\label{large-level-asymptotics}

The Hermite quantization equation shows numerically that the roots
approach an almost arithmetic sequence. Earlier analyses of the
fractional-power charts obtained the effective-level law
\(a_{n} \simeq n + 1/3\) {[}25,26{]}. As shown below, the offset
\(\frac{1}{3}\) is the Maslov index \(\gamma_{M}\) in the convention of
Ref. {[}38{]}. The power-Liouville representation makes it convenient to
refine this result through the \(t = 3\) chart, where the quartic
eigencoupling becomes an ordinary energy.

\subsection{Langer action and Maslov
index\label{langer-action-and-maslov-index}}

Write the reduced \(t = 3\) equation in the form

\begin{equation}
\left\lbrack - \frac{d^{2}}{dx^{2}} + \frac{\nu^{2} - 1/4}{x^{2}} + Ax^{2/3} \right\rbrack\chi(x)\mathcal{= E}\chi(x),
\end{equation}

with

\begin{equation}
\nu = \frac{5}{3},\quad\quad\mathcal{E}=\frac{2m\mathcal{E}_{3}}{\hslash^{2}},\quad\quad A = \frac{2mV_{6}L^{16/3}}{9\hslash^{2}} > 0.
\end{equation}

Here \(\mathcal{E}\) is the reduced \(t = 3\) energy and \(A > 0\) is
the fractional-power confinement coefficient.

Following the radial Langer/Maslov structure for power-law potentials
{[}38,39{]}, the action is

\begin{equation}
I\left( \mathcal{E} \right) = \int_{x_{in}}^{x_{out}}\sqrt{\mathcal{E}-Ax^{2/3} - \frac{\nu^{2}}{x^{2}}}\, dx,
\end{equation}

where \(x_{in}\) and \(x_{out}\) are the two positive turning points of
the Langer-modified momentum.

Introduce

\begin{equation}
a = \frac{3\mathcal{E}^{2}}{16A^{3/2}}.
\end{equation}

Here a is the same invariant Hermite-order parameter introduced in Eq.
(48); no second quantity is being introduced in the semiclassical
analysis.

After introducing the dimensionless coordinate \(y\) through
\(x = \left( \mathcal{E/}A \right)^{3/2}y\), let \(J(\lambda)\) denote
the scaled action function and \(\lambda\) the dimensionless endpoint
parameter. The action becomes

\begin{equation}
I = \frac{16a}{3}J(\lambda),
\end{equation}

Explicitly, the scaled action function is

\begin{equation}
J(\lambda) = \int_{y_{in}}^{y_{out}}\sqrt{1 - y^{2/3} - \frac{\lambda}{y^{2}}}\, dy,
\end{equation}

where its integration limits are the scaled turning points corresponding
to those in Eq. (69), while the dimensionless endpoint parameter is

\begin{equation}
\lambda = \left( \frac{3\nu}{16a} \right)^{2}.
\end{equation}

A matched expansion at the singular lower endpoint gives, as
\(\lambda \rightarrow 0\),

\begin{equation}
J(\lambda) = \frac{3\pi}{16} - \frac{\pi}{2}\lambda^{1/2} - \frac{3}{20}B\left( \frac{1}{6}, - \frac{1}{2} \right)\lambda^{5/6} + o\left( \lambda^{5/6} \right).
\end{equation}

The first term is the ordinary \(x^{2/3}\) action; the second is the
Langer endpoint contribution. Together with the half-integer
turning-point term in Eq. (77), it fixes the net quantum-number offset
identified below with the Maslov index. The third term is the leading
algebraic interaction between the centrifugal boundary layer and the
fractional-power confinement. Here \(B(p,q)\) denotes the Euler beta
function, understood by analytic continuation where required. Details
are given in Appendix A.

It follows that

\begin{equation}
\frac{I}{\pi} = a - \frac{\nu}{2} - C_{\nu}a^{- 2/3} + o\left( a^{- 2/3} \right),
\end{equation}

where

\begin{equation}
C_{\nu} = \frac{B(1/6, - 1/2)}{40\pi}\left( \frac{3\nu}{2} \right)^{5/3}.
\end{equation}

The Langer quantization rule

\begin{equation}
I = \pi\left( k + \frac{1}{2} \right),\quad\quad k = 0,1,2,\ldots,
\end{equation}

therefore gives

\begin{equation}
a_{k} = K_{k} + C_{\nu}K_{k}^{- 2/3} + o\left( K_{k}^{- 2/3} \right),
\end{equation}

with

\begin{equation}
K_{k} = k + \frac{\nu + 1}{2}.
\end{equation}

For the present \(N = 4\) orbit, \(\nu = 5/3\) and the physical root
label is \(n = k + 1\). Reexpressing the Langer result in terms of this
physical label, the constant quantum-number offset is the Maslov index
in the convention of Ref. {[}38{]}; explicitly,
\(\gamma_{M} = \frac{\nu - 1}{2} = \frac{1}{3}\). Hence

\begin{equation}
a_{n} = n + \gamma_{M} + C_{4}\left( n + \gamma_{M} \right)^{- 2/3} + o\left( n^{- 2/3} \right),
\end{equation}

where

\begin{equation}
C_{4} = \frac{B(1/6, - 1/2)}{40\pi}\left( \frac{5}{2} \right)^{5/3} = 0.1779993954\ldots.
\end{equation}

The corresponding quartic eigencouplings are

\begin{equation}
V_{4,n} \sim - 4\left( \frac{\hslash^{2}}{2m} \right)^{1/4}V_{6}^{3/4}\left\lbrack n + \gamma_{M} + C_{4}\left( n + \gamma_{M} \right)^{- 2/3} \right\rbrack^{1/2}.
\end{equation}

We deliberately stop at this order. Beyond the leading Langer structure,
higher-order radial Wentzel--Kramers--Brillouin (WKB) corrections
require more care than a direct one-dimensional Dunham substitution, and
no additional term is needed for the present spectral identification.

\subsection{Accuracy}\label{accuracy}

The correction is effective already at \(n = 10\): using the exact root
in Table 1, the leading approximation gives \(10.333333333333\ldots\)
and the corrected expression gives \(10.370852936332\ldots\). The
relative error falls from \(3.23 \times 10^{- 3}\) to
\(3.89 \times 10^{- 4}\).

At \(n = 20\) the relative error falls from about
\(1.1 \times 10^{- 3}\) to \(1.0 \times 10^{- 4}\), and by \(n = 30\) it
is about \(4.6 \times 10^{- 5}\). Over the range \(10 \leq n \leq 30\),
the remaining absolute difference between the exact root and the
corrected approximation is empirically close to being proportional to
\(n^{- 1}\). We use this observation only as a diagnostic of the next
unresolved order and do not identify it with any specific higher-order
radial WKB term. The next term of the matched-action expansion would
enter the root sequence at the same inverse-level order at which terms
beyond the leading Langer treatment can also contribute; no coefficient
is therefore assigned at this order. Figure 1 shows the improvement over
the first thirty physical roots.

For the levels shown, the corrected approximation lies slightly above
the exact roots. No monotone convergence is inferred, because the next
endpoint contribution and higher-order radial-WKB terms lie outside the
present leading-Langer treatment.

\begin{figure}
\centering
\includegraphics[width=\linewidth]{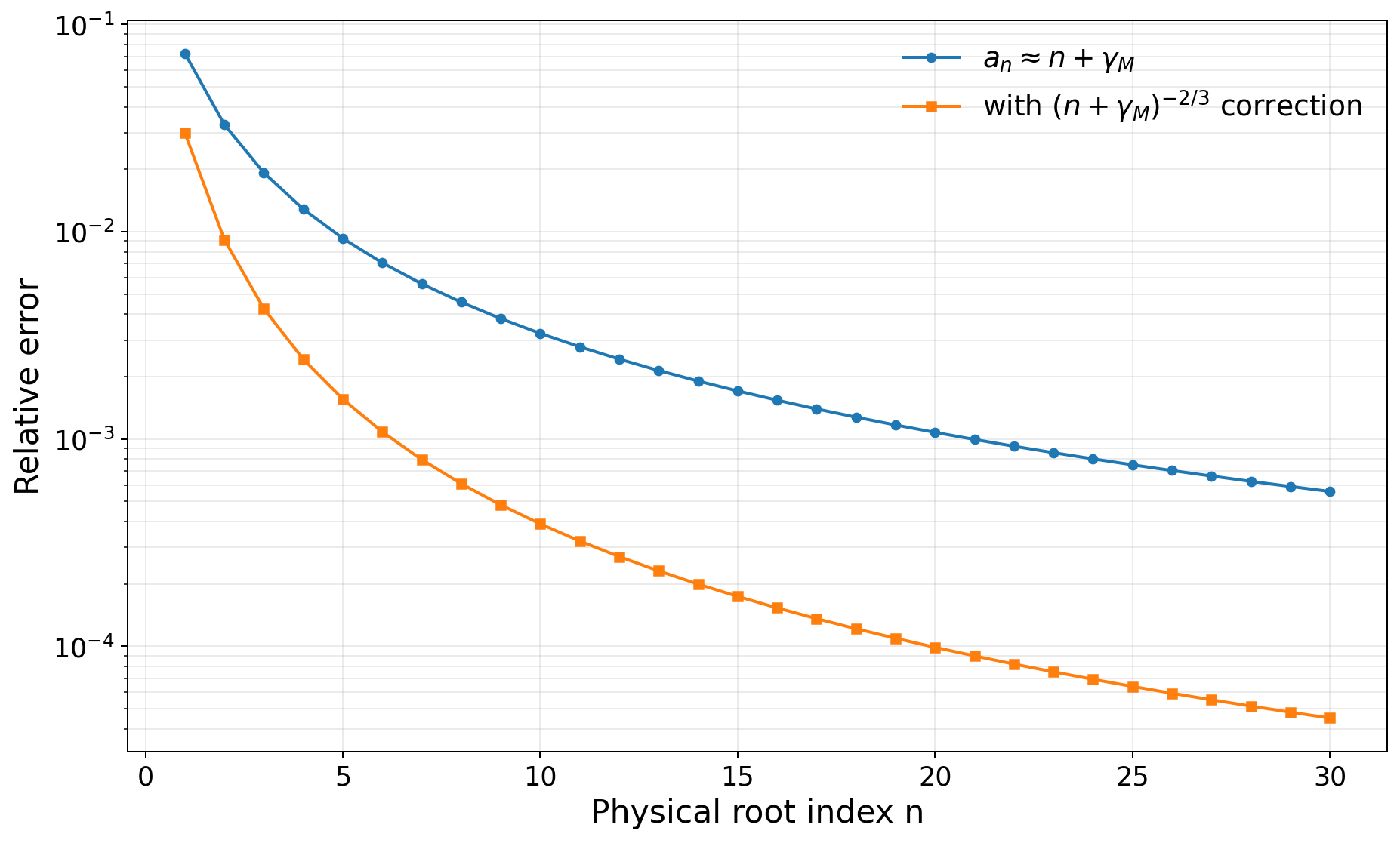}
\caption{Relative error of the leading and corrected large-n formulas
for the exact physical roots of the common Hermite spectrum equation.}
\end{figure}

\section{Discussion and
conclusions\label{discussion-and-conclusions}}

We have identified an exact four-member power-Liouville Hermite orbit
centered on the \(N = 4\) member of the finite-Hermite singular sextic
hierarchy. The orbit contains the selected-energy singular sextic
oscillator itself, a singular linear-plus-quadratic half-line
oscillator, the conditionally integrable \(x^{2/3}\) potential, and the
conditionally integrable inverse-square-root potential.

The structural invariant is the Frobenius index. For the \(N = 4\)
sextic problem it equals five, and under \(x \propto r^{t}\) it becomes
\(5/t\). This immediately produces the otherwise rather special-looking
centrifugal coefficients \(91/36\) and \(21/16\) in reduced units, or
\(91\hslash^{2}/(72m)\) and \(21\hslash^{2}/(32m)\) in the physical
potentials. Both coefficients had previously arisen from separate
BHE/Hermite constructions {[}25,26{]}. The power-Liouville picture shows
that they are two projections of the same singular sextic sector. On the
selected physical slice of this member, the degree-five accessory
compatibility obstruction vanishes identically, so finite-Hermite
termination imposes no additional relation between the quartic and
sextic couplings. As noted in Sec. 3, identity cancellation also occurs
for a lower member; the distinctive feature of the present member is the
simultaneous match to both published fractional-power centrifugal
coefficients.

The central spectral statement is the exact invariance of the Wronskian
for a directly transported pair of regular and recessive solutions.
Independently normalized spectral functions may differ by
parameter-dependent factors, but away from normalization-degeneracy
points they have the same physical zero set. The \(t = 3\) and \(t = 4\)
systems are already a standard direct power-duality pair, so their
energy-coupling exchange does not require the sextic anchor. What the
present four-member orbit adds is their embedding into the \(N = 4\)
finite-Hermite sextic sector and the transport of the same physical zero
to the \(t = 1\) and \(t = 2\) charts. The explicit Hermite identity
\(F_{3} = - F_{4}\) then gives a particularly transparent representation
of this common zero condition.

This interpretation clarifies the meaning of the ``privileged''
couplings of the singular sextic oscillator. In the \(t = 1\) chart, the
energy is fixed at \(E = V_{0}\) and the quartic coefficient is
quantized. In the \(t = 3\) chart, the same parameter becomes the
ordinary bound-state energy. The exact root sequence \(a_{n}\) therefore
gives the quartic eigencouplings in Eq. (62).

Because the node count and Wronskian zero are preserved, they provide
chart-independent physical labels; the orbit is therefore not merely a
formal mapping between solvable differential equations.

Finally, the \(t = 3\) chart provides a convenient semiclassical
description. The leading effective-level law
\(a_{n} \simeq n + \gamma_{M}\), with \(\gamma_{M} = \frac{1}{3}\),
acquires a first matched-endpoint correction within the leading Langer
action, proportional to \(\left( n + \gamma_{M} \right)^{- 2/3}\). This
correction is not inserted phenomenologically within that leading-action
approximation: it follows from a matched expansion of the Langer action
at the singular endpoint. Its substantial numerical improvement provides
an independent check on the orbit-based spectral identification.

The chart equivalence established here should not be confused with
strict Darboux isospectrality of fixed Hamiltonians, such as the higher
fractional-power BHE branch studied in Ref. {[}29{]}. The present result
instead concerns a common physical zero set transported through an
energy-coupling reparametrization.

A natural continuation is to study the algebraic compatibility loci at
other hierarchy indices, including the lower identity-cancellation
member, and to determine which of them carry useful physical Sturmian
spectra or further power-Liouville connections.

\section{Appendix A. Matched expansion of the fractional-power
action}\label{appendix-a.-matched-expansion-of-the-fractional-power-action}

We give the cutoff matching that produces the matched-endpoint
correction used in Sec. 7. Consider

\begin{equation}
J(\lambda) = \int_{y_{in}}^{y_{out}}\sqrt{1 - y^{2/3} - \lambda y^{- 2}}\, dy,\quad\quad\lambda \rightarrow 0.
\end{equation}

At \(\lambda = 0\),

\begin{equation}
J(0) = \int_{0}^{1}\sqrt{1 - y^{2/3}}\, dy = \frac{3}{2}B\left( \frac{3}{2},\frac{3}{2} \right) = \frac{3\pi}{16}.
\end{equation}

A regular Taylor expansion in \(\lambda\) is not uniform because the
derivative with respect to \(\lambda\) diverges at the lower endpoint.
Choose an overlap cutoff \(\eta = \lambda^{p}\), where \(p\) is the
matching exponent and \(5/14 < p < 7/18\), so that
\(\sqrt{\lambda} \ll \eta \ll 1\), and split the action into inner and
outer pieces:

\begin{equation}
J(\lambda) = J_{in}(\lambda,\eta) + J_{out}(\lambda,\eta),\quad\quad J_{in} = \int_{y_{in}}^{\eta}(\cdots)\, dy,\quad J_{out} = \int_{\eta}^{y_{out}}(\cdots)\, dy.
\end{equation}

In the outer region, expansion at fixed \(y\) gives

\begin{equation}
J_{out} = J(0) - \eta + \frac{3}{10}\eta^{5/3} - \frac{\lambda}{2\eta} - \frac{3\lambda}{4}\eta^{- 1/3}.
\end{equation}

The omitted outer terms are
\(O\left( \eta^{7/3},\lambda,\lambda^{2}\eta^{- 3} \right)\). With the
overlap window just specified, all three are asymptotically smaller than
the retained \(\lambda^{5/6}\) contribution under the stated overlap
scaling.

For the inner region set \(y = \sqrt{\lambda}\, u\) and
\(U = \eta/\sqrt{\lambda}\). The lower turning point is
\(u_{in} = 1 + O\left( \lambda^{1/3} \right)\), and replacing it by 1
changes the action only beyond the order retained. The first omitted
inner contribution has the same order as the first outer remainder,
while the subsequent large-U remainders are smaller under the same
overlap window. One obtains

\begin{equation}
J_{in} = \sqrt{\lambda}\int_{1}^{U}\sqrt{1 - u^{- 2}}\, du - \frac{\lambda^{5/6}}{2}\int_{1}^{U}\frac{u^{2/3}}{\sqrt{1 - u^{- 2}}}\, du + o\left( \lambda^{5/6} \right).
\end{equation}

Here FP is the Hadamard finite part, equivalently the analytic
continuation of the corresponding Euler beta integral after subtraction
of the divergent endpoint terms. The second integral is

\begin{equation}
FP\int_{1}^{\infty}\frac{u^{5/3}}{\sqrt{u^{2} - 1}}\, du = \frac{1}{2}B\left( - \frac{5}{6},\frac{1}{2} \right) = \frac{3}{10}B\left( \frac{1}{6}, - \frac{1}{2} \right).
\end{equation}

Define
\(C_{B} = \frac{3}{20}B\left( \frac{1}{6}, - \frac{1}{2} \right)\).
Using the large-\(U\) expansions of the two inner integrals then gives

\begin{equation}
J_{in} = \eta - \frac{\pi}{2}\lambda^{1/2} - \frac{3}{10}\eta^{5/3} + \frac{\lambda}{2\eta} + \frac{3\lambda}{4}\eta^{- 1/3} - C_{B}\lambda^{5/6} + o\left( \lambda^{5/6} \right).
\end{equation}

Adding Eqs. (86) and (89) cancels the cutoff-dependent terms \(\eta\),
\(\eta^{5/3}\), \(\lambda/\eta\), and \(\lambda\eta^{- 1/3}\), leaving

\begin{equation}
J(\lambda) = \frac{3\pi}{16} - \frac{\pi}{2}\lambda^{1/2} - \frac{3}{20}B\left( \frac{1}{6}, - \frac{1}{2} \right)\lambda^{5/6} + o\left( \lambda^{5/6} \right).
\end{equation}

Using Eqs. (71) and (73), Eq. (90) yields

\begin{equation}
\frac{I}{\pi} = a - \frac{\nu}{2} - \frac{B(1/6, - 1/2)}{40\pi}\left( \frac{3\nu}{2} \right)^{5/3}a^{- 2/3} + o\left( a^{- 2/3} \right).
\end{equation}

Equation (91) is the result used in the main text. It is the first
nonanalytic correction generated within the leading Langer action.
Higher-order radial WKB terms require a separate uniform treatment and
are not included in the present expansion. Accordingly, we do not
identify this term as the first correction to the exact roots without
qualification; throughout the paper it is described as the first
matched-endpoint correction within the leading Langer action.

\section{Appendix B. Explicit Schr\"{o}dinger-to-BHE map and compatibility
condition}\label{appendix-b.-explicit-schruxf6dinger-to-bhe-map-and-compatibility-condition}

To make the construction used in Sec. 3 self-contained, we record the
transformation and the local compatibility condition explicitly.
Following Ref. {[}13{]}, set

\begin{equation}
z = \frac{r^{2}}{4},\quad\quad\psi(r) = z^{\alpha_{0}}\exp\left( \alpha_{1}z + \alpha_{2}z^{2} \right)u(z)
\end{equation}

with

\begin{equation}
\left( \alpha_{0},\alpha_{1},\alpha_{2} \right) = \left( \frac{2\gamma - 1}{4},\frac{\delta}{2},\frac{\varepsilon}{4} \right)
\end{equation}

Then the singular sextic Schr\"{o}dinger equation is transformed into the
canonical BHE (19) when

\begin{equation}
\gamma = 1 \pm \frac{1}{2}\sqrt{1 + \frac{8mV_{- 2}}{\hslash^{2}}},\quad\quad\delta = \frac{64mV_{4}}{\hslash^{2}\varepsilon},\quad\quad\varepsilon = \pm 16\sqrt{\frac{2mV_{6}}{\hslash^{2}}}
\end{equation}

and

\begin{equation}
\alpha = - \frac{8mV_{2}}{\hslash^{2}} + \frac{\delta^{2}}{4} + \frac{\gamma + 1}{2}\varepsilon,\quad\quad q = - \frac{\gamma\delta}{2} - \frac{2m\left( E - V_{0} \right)}{\hslash^{2}}
\end{equation}

Choosing the terminating branch \(\gamma = - N\) and imposing
\(E = V_{0}\) and \(V_{2} = 0\) gives Eq. (29). The signs in \(\gamma\)
and \(\varepsilon\) select the fundamental representation; for the
physical confined sextic branch the sign of the quartic-Gaussian factor
is chosen so that the wave function is recessive at infinity.

The local compatibility polynomial follows from a Frobenius series at
the regular singular point,

\begin{equation}
u(z) = \sum_{n = 0}^{\infty}c_{n}z^{n},\quad\quad c_{- 1} = 0
\end{equation}

Substitution into Eq. (19) gives

\begin{equation}
(n + 1)(n + \gamma)c_{n + 1} + (\delta n - q)c_{n} + \left\lbrack \alpha + \varepsilon(n - 1) \right\rbrack c_{n - 1} = 0
\end{equation}

For \(\gamma = - N\) and Eq. (29), Eq. (97) reduces to Eq. (30). At
\(n = N\), the coefficient of \(c_{N + 1}\) vanishes; consistency of the
first \(N + 1\) equations therefore yields the accessory compatibility
polynomial \(\mathcal{R}_{N} = D_{N}\). This clarifies why the
coefficients \(c_{n}\) in Sec. 3 are Frobenius coefficients rather than
the coefficients of the Hermite expansion.

For \(\varepsilon \neq 0\), the same BHE solution admits the Hermite
expansion used in Ref. {[}13{]},

\begin{equation}
u(z) = \sum_{n = 0}^{N}b_{n}H_{\nu_{0} + n}(\xi),\quad\quad\nu_{0} = \gamma - \frac{\alpha}{\varepsilon},\quad\quad\xi = s_{0}\left( z + \frac{\delta}{\varepsilon} \right),\quad s_{0} = \pm \sqrt{- \frac{\varepsilon}{2}}
\end{equation}

The Hermite series terminates when \(\gamma = - N\) and the same
degree-\(N + 1\) accessory condition is satisfied. By the three-term
recurrence of the Hermite functions, every such finite sum can be
reduced to two contiguous Hermite functions. In particular, for
\(N = 4\) one may write

\begin{equation}
u(z) = P_{0}(z)H_{- \alpha/\varepsilon}(\xi) + P_{1}(z)H_{- 1 - \alpha/\varepsilon}(\xi),\quad\quad\deg P_{0} = 2,\quad\deg P_{1} = 3
\end{equation}

This gives the solution-level finite-Hermite realization underlying the
spectral construction. The explicit polynomial coefficients are fixed by
the recurrence but are not needed for the Wronskian-zero argument of
Sec. 5.

%
%
%


\end{document}